\documentclass[11pt,letterpaper]{article}

\usepackage[T1]{fontenc}
\usepackage[utf8]{inputenc}
\usepackage{lmodern}
\usepackage[margin=1in]{geometry}
\usepackage{amsmath,amssymb,amsthm,mathtools}
\usepackage{graphicx,xcolor,booktabs,array,tabularx,placeins}
\usepackage{microtype}
\usepackage[numbers,sort&compress]{natbib}
\usepackage{xurl}
\usepackage{authblk}
\usepackage[font=small,labelfont=bf]{caption}
\usepackage{hyperref}
\hypersetup{colorlinks=true,linkcolor=blue!45!black,citecolor=blue!45!black,urlcolor=blue!45!black}
\providecommand{\Description}[2][]{}

\theoremstyle{plain}
\newtheorem{theorem}{Theorem}[section]
\newtheorem{lemma}[theorem]{Lemma}
\newtheorem{proposition}[theorem]{Proposition}

\theoremstyle{definition}

\theoremstyle{plain}

\newcommand{\E}{\mathbb E}
\newcommand{\Prb}{\mathbb P}
\newcommand{\OPT}{\mathrm{OPT}}

\newcommand{\st}{\mathrm{st}}
\newcommand{\unif}{\mathrm{unif}}
\newcommand{\dyn}{\mathrm{dyn}}

\title{Pricing Information in Search with Nonobligatory Inspection}
\date{}
\hypersetup{pdftitle={Pricing Information in Search with Nonobligatory Inspection}}

\author{Shuo Zhang, Boyu Liu, and Zihe Wang}

\affil{%
 Renmin University of China\\
  \{zhangshuo1422, liuboyu03, wang.zihe\}@ruc.edu.cn
}

\date{}

\hypersetup{
  pdfauthor={Shuo Zhang, Boyu Liu, Zihe Wang}
}

\begin{document}
\raggedbottom
\maketitle
\begin{abstract}
As AI agents reshape Web search and advertising, platforms can monetize information that improves decisions. We study information pricing for sequential search among independent alternatives with nonobligatory inspection: the buyer may select any alternative without buying its report. We show that a common fee maximizes static revenue for identical priors, resolving a uniform-pricing conjecture. Our results also establish that exact optimization is NP-hard for heterogeneous binary priors. Despite this hardness, our polynomial-time static pricing algorithm achieves a $3$-approximation to the information surplus. More generally, a common static fee approximates optimal adaptive revenue within a logarithmic factor in the number of alternatives. This guarantee is asymptotically tight.

\end{abstract}

\section{Introduction}

Web platforms have long monetized product discovery through sponsored placement and user clicks~\cite{edelman2007,athey2011}. This model draws value from limited human attention: display position matters because users can examine only a small set of alternatives. Agent-assisted shopping is already available on major platforms (e.g., Google and Amazon)~\cite{google2026travel,amazon2026shopping,paypal2025perplexity,walmart2026agents}. By gathering and comparing information on the user's behalf, agents can evaluate many more alternatives without requiring users to browse each one~\cite{rothschild2026economy,bansal2025marketplace,lucier2026agentic,shahidi2026coase,rothschild2026search}. This can make prominent advertising slots less central to product discovery and weaken the connection between user clicks and the value of search~\cite{rothschild2026economy,rothschild2026search}. These changes motivate a complementary revenue model: selling information that improves the agent's choice. Recent work explores subsidized inspection in agent-mediated markets~\cite{candelas2026} and marketplaces for publisher information used in AI answers~\cite{mobius2026information}.

Consider a buyer, or an agent acting on the buyer's behalf, comparing suppliers using public listings. The platform sells reports on independent alternatives, each revealing one alternative's value. The buyer can purchase reports sequentially and still select a supplier whose report was not purchased. This is search with \emph{nonobligatory inspection}~\cite{doval2018,beyhaghi2019}, in contrast to the classical Pandora's box model, where selection requires inspection~\cite{weitzman1979}. The buyer chooses which reports to purchase and when to stop to maximize the expected value of the selected alternative net of report payments. Our focus is on \emph{static pricing}: the seller commits to a fee for each report before search, and these fees remain fixed as the buyer learns.

Nonobligatory inspection creates a tension between selling information and preserving the buyer's fallback. An uninspected alternative offers its prior mean, but a disappointing report replaces this prospect with a low realized value. The seller must therefore account for how one inspection changes the demand for other reports and the option of ending search without further purchases. The last report illustrates the difficulty: it can be worthless after either very good earlier outcomes, which make an inspected alternative attractive, or very bad ones, which favor selecting the remaining alternative blindly. Even identical priors can thus generate different willingness to pay as search unfolds.

\subsection{Results and approach}

We measure revenue against the \emph{information surplus} $\Delta$, the improvement in expected choice value from learning all values rather than choosing without inspection. The buyer can secure the latter payoff by declining every report, so this surplus bounds revenue under any pricing rule. A valuable alternative may generate little demand for its report if the buyer would choose it anyway; what matters for revenue is whether information can change the choice. Table~\ref{tab:results} summarizes our results on uniform optimality, approximation under heterogeneity, and the revenue gain from adaptive pricing.

\begin{table}[t]
\centering
\caption{Summary of our main results for $n$ independent alternatives.}
\label{tab:results}
\small
\setlength{\tabcolsep}{6pt}
\renewcommand{\arraystretch}{1.12}
\begin{tabularx}{\textwidth}{@{}>{\raggedright\arraybackslash}p{.25\textwidth}>{\raggedright\arraybackslash}X@{}}
\toprule
\textbf{Prior structure} & \textbf{Main results} \\
\midrule
Identical priors & A common fee attains optimal static revenue (Theorem~\ref{thm:iid:uniform}). \\
\addlinespace[4pt]
Heterogeneous binary priors & Exact static optimization is NP-hard (Theorem~\ref{thm:binary:hardness}).\newline A polynomial-time static scheme earns at least one third of the information surplus (Theorem~\ref{thm:binary-three}). \\
\addlinespace[4pt]
General independent priors & A common fee gives a $(1+\ln n)$-approximation to optimal adaptive revenue.\newline The worst-case adaptive-to-static revenue ratio is ${(1+o(1))\ln n}$ (Theorem~\ref{gen:adaptivegap}). \\
\bottomrule
\end{tabularx}
\end{table}

We show that a common fee maximizes static revenue for any number of alternatives with identical priors (Theorem~\ref{thm:iid:uniform}), resolving the uniform-pricing conjecture in~\cite{chawla2026}. Although reports are initially identical, symmetry alone does not imply that they should carry the same price: a cheap report could induce search and create demand for more expensive ones. Our proof rules out this possibility by simplifying an arbitrary fee vector. The buyer has an optimal response that inspects cheaper reports first, giving an exact revenue expression with a correction for the final report. Crucially, this correction depends only on the largest fee. Adjacent price groups can then be merged by either raising the lower fee or lowering the higher one; at least one of these changes does not reduce revenue. Repeating the argument yields a common fee. An optimal seller therefore need not withhold a report or discriminate across reports, even though the buyer's willingness to pay changes during search.

Once priors differ, the seller must decide which alternatives to preserve as blind choices as well as how much to charge. We prove that finding optimal static prices is NP-hard even for binary (two-point) priors (Theorem~\ref{thm:binary:hardness}). The reduction encodes integers in low-outcome probabilities and uses prices to place reports before or after inspection of the best blind alternative. Independence turns the relevant event probabilities into products, so finding optimal prices solves a Product Partition instance.

Despite this hardness, our polynomial-time algorithm achieves a $3$-approximation to the information surplus (Theorem~\ref{thm:binary-three}). The idea is to separate the surplus into the gain from searching while the best blind alternative is preserved and the additional gain from inspecting it. One scheme withholds its report and extracts the first component. Two further schemes address the remaining value by either selling this report early, while another blind choice is available, or revealing a competing value for free to encourage its purchase. A binary report's value peaks when the competing value equals the reported alternative's prior mean. This shape gives revenue guarantees that together cover the surplus. Choosing the largest guarantee therefore yields revenue of at least $\Delta/3$, without optimizing the entire search process. The guarantee is independent of the number of alternatives and the extent of heterogeneity; whether static prices can guarantee a larger share remains open.

More generally, we prove that a common static fee earns at least a ${1/(1+\ln n)}$ share of optimal adaptive revenue, where $n$ is the number of alternatives (Theorem~\ref{gen:adaptivegap}). The guarantee is asymptotically tight. Adaptive prices here may respond to earlier inspections under a rule announced before search. Our analysis uses aggregate demand for reports: the area under the demand curve equals the information surplus, while revenue at a common fee is the area of a rectangle beneath it. The limit of $n$ purchases lets us bound this area by a logarithmic multiple of the largest rectangle, without tracing the buyer's choices at every possible history.

An adaptive seller can collect the full surplus by charging $\Delta$ for the first report and making all subsequent reports free (Proposition~\ref{gen:dynamicexact}). Knowing this rule in advance, the buyer values the first purchase as access to full information, and the payment leaves the initial blind payoff. Thus even a rule that only tracks whether a purchase has occurred attains optimal adaptive revenue. The matching lower bound comes from identical exponential priors: as the number of alternatives grows, the information surplus grows logarithmically while optimal static revenue stays bounded. Per-report fees lose future sales once the buyer finds a sufficiently good alternative, whereas an access fee collects the value of information before search begins.

\subsection{Related work}

Our problem connects sequential search with the sale of information. Pandora's box models costly inspection~\cite{weitzman1979,beyhaghi2023survey}, with extensions to combinatorial constraints, online arrivals, and correlated values~\cite{singla2018,esfandiari2019,chawla2020correlations,chawla2023correlations,gergatsouli2023}. Allowing selection without inspection changes the structure and computational complexity of search~\cite{doval2018,beyhaghi2019,fu2023,beyhaghi2023,scully2024}. Related work examines how auctions coordinate information acquisition~\cite{kleinberg2016} and how product prices, prominence, and search order influence consumer behavior~\cite{armstrong2009,armstrong2017,gamp2022}. Our focus is on pricing the information that guides this search.

The information-selling literature studies disclosure protocols, menus of experiments, and consulting mechanisms~\cite{babaioff2012,bergemann2018,chen2020}, alongside markets for consumer information and query pricing~\cite{bergemann2015,bergemann2019,koutris2013,koutris2015,chawla2019query}. In our setting, each report reveals one value, and its price affects both sequential learning and the choice of uninspected alternatives. With nonobligatory inspection, \citet{chawla2026} establish uniform optimality for two i.i.d. alternatives and for any number of i.i.d. binary alternatives. For general i.i.d. priors, they obtain a $(1-1/n)$ share of optimal static revenue by withholding one report and posting a common fee for the rest. Our uniform-optimality result removes this loss while keeping every report available. For heterogeneous binary priors, they obtain $\Delta/2$ through free disclosure followed by contingent pricing. Our results also establish the computational difficulty of exact static pricing and give revenue guarantees with all fees fixed before disclosure.

\noindent\textbf{Organization.}\enspace
Section~\ref{sec:model} introduces the model. Sections~\ref{sec:iid}--\ref{gen:section} study uniform optimality, binary pricing, and the static--adaptive gap. Section~\ref{sec:conclusion} discusses open questions, and the appendices provide the remaining proofs.

\section{Model and benchmarks}
\label{sec:model}

A seller offers reports on $n\ge2$ independent alternatives. Alternative $i$ has a nonnegative value $X_i$ and a finite prior mean $\mu_i=\E[X_i]$. Their prior distributions are common knowledge, but neither party initially observes the realizations. Under static pricing, the seller posts a fixed fee $p_i$ for report $i$ before search. Paying this fee reveals the value $X_i$. Reports have no production costs; static fees are charged separately, without access fees or bundles. A zero fee provides the report for free; an infinite fee withholds it but leaves the alternative available for selection. Section~\ref{gen:section} also considers adaptive fees under a rule announced before search.

The risk-neutral buyer purchases reports sequentially and may stop to select an alternative or take a payoff of zero. An uninspected alternative remains available as a \emph{blind choice} worth its prior mean. Independence ensures that learning about other alternatives does not change this expected value. Denote the set of uninspected alternatives by $S$ and the best value revealed so far by $b$, initially zero. The expected payoff from stopping is
\begin{equation}
V(S,b)=\max\{b,\max_{i\in S}\mu_i\}.
\label{eq:model:stop}
\end{equation}
An empty maximum is zero. Inspection replaces a prior mean with a realized value, changing this stopping payoff. Past report payments are sunk when deciding whether to continue.

The buyer maximizes expected utility: the expected value of the selected alternative minus expected total payments. Each purchase is evaluated in light of how its outcome may affect later search. Among policies that give the same utility, ties favor the seller's expected revenue (Appendix~\ref{app:model-details}). For the static fee vector $\mathbf p=(p_1,\ldots,p_n)$, expected total payments give the seller's revenue $R(\mathbf p)$. We denote optimal revenue from arbitrary static fees by $\OPT_{\st}$ and from a common fee by $\OPT_{\unif}$; both benchmarks are defined as suprema.

To bound these revenues, consider the buyer's best choice without information. Selecting the alternative with the highest prior mean gives the blind payoff $\mu_* = \max_i\mu_i$. Full information instead allows selection of the best realized value. The gain in expected choice value is the \emph{information surplus} $\Delta$:
\begin{equation}
\Delta=\E[\max_i X_i]-\mu_*.
\label{eq:model:surplus}
\end{equation}
The buyer can always decline every report and keep the blind payoff. Only the improvement over this outside option can therefore become revenue, giving $0\le\OPT_{\unif}\le\OPT_{\st}\le\Delta$.

The surplus bounds total revenue, but static pricing also limits what the buyer will pay for any single report. The buyer can leave that alternative uninspected and retain its prior mean as a fallback. This comparison gives a cap that holds throughout search. The positive part of $x$ is $x^+=\max\{x,0\}$.
\begin{lemma}[Individual information cap]
\label{gen:cap}
Define the information cap $c_i=\E[(X_i-\mu_i)^+]$. Under a static fee vector, at any history where $i$ is uninspected, no utility-optimal continuation buys its report at a fee strictly above $c_i$. Replacing such a fee by $c_i$ preserves optimal buyer utility and weakly increases seller-favorable revenue. All fees can be capped successively.
\end{lemma}

The cap measures potential gains above the prior mean. Imagine keeping the alternative blind as a fallback while searching elsewhere. Since the buyer selects only one alternative, raising this fallback by one unit improves continuation utility by at most one unit. A high realization can therefore add at most its excess over the mean; a low one cannot improve on keeping the alternative blind. Averaging gives the cap $c_i$, even when inspection changes later search. Lowering an excessive fee to this cap can induce purchases without sacrificing revenue. A deterministic alternative thus has no information to sell, however valuable it is to choose. Appendix~\ref{app:cap-proof} gives the proof.

\section{Uniform pricing}
\label{sec:iid}

The information cap still allows fees to differ across reports. With identical priors, a seller might use a cheap report to encourage search and create demand for expensive reports later. Our first result shows that this cannot outperform a common fee. We characterize the buyer's optimal response, express revenue through purchase probabilities, and use this expression to merge fees without losing revenue.

\begin{theorem}[Optimal uniform static pricing]
\label{thm:iid:uniform}
For $n\ge2$ independent copies of any nonnegative finite-mean value, every static fee vector is weakly dominated in revenue by a common fee. An optimal common fee is attained.
\end{theorem}

The difficulty is that willingness to pay changes during search, even though the priors are identical. Consider two independent alternatives, each equally likely to be worth $0$, $5$, or $10$, with both reports priced at $1$. Buying one report is better than choosing blindly: even without further inspection, it raises expected choice value from $5$ to $20/3$, more than covering its fee. If the first value is $0$, the buyer selects the remaining alternative blindly; if it is $10$, the buyer keeps the revealed value. If it is $5$, buying the final report raises expected choice value to $20/3$. This gain of $5/3$ exceeds the fee, so the buyer inspects again.

\begin{figure}[t]
\centering
\includegraphics[width=0.62\linewidth]{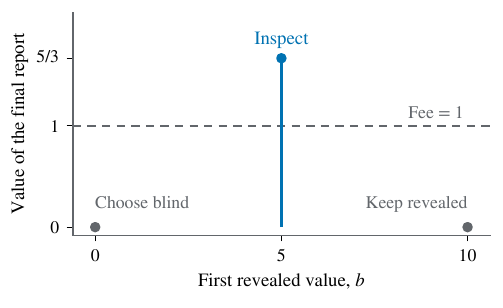}
\caption{An intermediate observation makes the final report worth buying. Each of two alternatives independently takes values $0$, $5$, and $10$ with equal probability, and each report costs $1$. When the first report reveals $b=0$, $5$, or $10$, the last report is worth $0$, $5/3$, or $0$, respectively. Only $b=5$ leads to further inspection.}
\Description{Three points show the value of the final report after the first inspection reveals zero, five, or ten. These values are zero, five thirds, and zero. A horizontal line marks the fee of one. The buyer chooses the remaining alternative blindly after observing zero, inspects it after observing five, and keeps the inspected alternative after observing ten.}
\label{fig:report-value}
\end{figure}

Figure~\ref{fig:report-value} shows all three possible histories. Both very good and very poor earlier draws can end search, while an intermediate value sustains demand for the final report. Earlier inspections are different because another uninspected alternative still supplies the common mean as a fallback. This distinction shapes the buyer's optimal stopping rule and, in turn, the seller's revenue.

\subsection{Buyer response and revenue}

Let $X$ be a value drawn from the common prior, with mean $\mu$, cumulative distribution function $F$, and information cap ${c=\E[(X-\mu)^+]}$. Constant values yield no revenue, so we focus on $c>0$ and cap all fees using Lemma~\ref{gen:cap}. An optimal buyer response can inspect cheaper reports first. Uninspected alternatives offer the same distribution and the same blind payoff, making their roles interchangeable. Paying the lower fee first saves money whenever search ends before the second purchase. The exchange argument below shows that this ordering preserves both optimal buyer utility and seller revenue.

\begin{samepage}
\begin{lemma}[Cheapest-first response]
\label{lem:iid:order}
At every history, there is a revenue-maximizing utility-optimal continuation that, whenever it inspects another alternative, inspects a cheapest remaining report. Equal-price ties may follow a fixed ordering of the labels.
\end{lemma}
\end{samepage}

\begin{proof}
Fix a history and a utility-optimal continuation that inspects report $j$ first. For another uninspected alternative $i$ with $p_i\le p_j$, let $\alpha$ be its probability of later inspection.

Interchange the roles of $i$ and $j$ throughout the continuation, including later inspections and terminal selection. Independence and identical priors preserve the distribution of the selected value and every other report's purchase probability. Expected payments for these two reports fall from $p_j+\alpha p_i$ to $p_i+\alpha p_j$, so utility improves by
$(p_j-p_i)(1-\alpha)\ge 0.$
Since the original continuation was optimal, this improvement must be zero. The exchange therefore preserves both utility and expected payments.

Start with a seller-favorable optimal continuation and repeat this exchange to place a cheapest report first. Equal fees allow a fixed label order. Applying the same argument at subsequent histories, by finite-horizon backward induction, gives the claimed response.
\end{proof}

With the order settled, revenue depends on when the buyer stops. A good revealed value can make further search unattractive, while a poor one can favor taking the last alternative blindly. For the cheapest remaining fee $p\in(0,c]$, the upper stopping threshold $g(p)$ and the lower threshold $\ell(p)$ solve
\begin{equation}
\mathbb E[(X-g(p))^+]=p,\qquad
\mathbb E[(\ell(p)-X)^+]=p.
\label{eq:iid:thresholds}
\end{equation}
The thresholds are unique and satisfy $\ell(p)\le\mu\le g(p)$. While at least two alternatives remain uninspected, an optimal response buys the cheapest report whenever the best revealed value is at most $g(p)$. Another blind alternative still supplies the mean $\mu$ as a fallback. Above $g(p)$, each remaining report costs more than its expected gain over the revealed value; combining purchases cannot help, since the buyer selects only one alternative. For the final report, inspection is optimal when the best revealed value lies in $[\ell(p),g(p)]$. Outside this interval, the buyer keeps either the blind choice or the best inspected alternative.

To calculate purchase probabilities, write $q(p)$ for the probability that a draw is at most the upper threshold and $z(p)$ for the probability that it falls below the lower threshold:
\begin{equation}
q(p)=F(g(p)),\qquad
z(p)=F(\ell(p)-)=\Prb[X<\ell(p)].
\label{eq:iid:qz}
\end{equation}
As the fee rises, $q$ falls and $z$ rises: fewer histories support a purchase. We set $q(0)=1$ and $z(0)=0$ for zero-fee terms.

\begin{proposition}[Exact revenue formula]
\label{prop:iid:revenue}
For sorted capped fees
$0\le p_1\le\cdots\le p_n\le c$,
\begin{equation}
R(\mathbf p)
=\sum_{j=1}^{n}p_jq(p_j)^{j-1}
-p_nz(p_n)^{n-1}.
\label{eq:iid:revenue-vector}
\end{equation}
In particular, the uniform-price revenue is
\begin{equation}
\mathcal R_n(p)
=p\left(\sum_{j=0}^{n-1}q(p)^j-z(p)^{n-1}\right).
\label{eq:iid:uniform-revenue}
\end{equation}
\end{proposition}

Each term in the sum is a fee times a purchase probability. Reaching report $j$ requires all earlier values to be at most its upper threshold $g(p_j)$. Since fees increase along the search order, upper thresholds decrease; this condition also ensures that the buyer chose to continue at every earlier step. Independence then gives the probability $q(p_j)^{j-1}$. For the final report, we subtract the histories where all earlier values fall below its lower threshold and the buyer takes the remaining alternative blindly. Appendix~\ref{app:revenue-proof} verifies the formula.

For the seller, the key feature is that the correction for blind selection depends only on the largest fee. A price change that improves the revenue sum without raising this fee also improves actual revenue. This connects the buyer's stopping behavior to the price comparison that follows.

\subsection{Merging prices without losing revenue}

Consider two adjacent groups of reports, charged a lower and a higher fee. Raising a fee earns more per purchase but reduces the chance of a sale. Later reports are more sensitive to this change because more preceding draws must fall below their stopping threshold. This difference in demand sensitivity lets us eliminate price differences: if raising the earlier group's fee loses revenue, lowering the later group's fee attracts enough purchases to offset the smaller payment. At least one merger preserves the revenue sum, and neither raises the largest fee. The correction for blind selection therefore cannot increase. We use this observation to adapt the price-merging argument of~\cite[Lemma~20]{chawla2026} to nonobligatory inspection.

\begin{proof}[Proof of Theorem~\ref{thm:iid:uniform}, except attainment]
First apply the individual-cap lemma (Lemma~\ref{gen:cap}), which preserves utility and weakly increases revenue, and sort the resulting prices. Write the revenue sum and the final-report correction as
\[
A(p_1,\ldots,p_n)
=\sum_{j=1}^{n}p_jq(p_j)^{j-1},
\qquad
D(t)=t z(t)^{n-1}.
\]
Then $R(\mathbf p)=A-D(p_n)$, and $D$ is nondecreasing.

Consider adjacent constant-price blocks $I$ and $J$, with the earlier block $I$ priced at $a$ and the later block $J$ priced at $b>a$. The aggregate continuation probabilities in the two blocks are
\[
A_I(t)=\sum_{j\in I}q(t)^{j-1},
\qquad
A_J(t)=\sum_{j\in J}q(t)^{j-1}.
\]
Raising block $I$ to $b$ changes $A$ by
$bA_I(b)-aA_I(a).$
Lowering block $J$ to $a$ changes $A$ by
$aA_J(a)-bA_J(b).$
At least one change is nonnegative. This is immediate for $a=0$. For $a>0$, put $q_a=q(a)\ge q_b=q(b)>0$. Every exponent in $J$ exceeds every exponent in $I$, so
\begin{equation}
\frac{A_J(a)}{A_J(b)}
\ge
\frac{A_I(a)}{A_I(b)}.
\label{eq:iid:block-ratios}
\end{equation}
Each ratio is a weighted average of $(q_a/q_b)^{j-1}$ with positive weights proportional to $q_b^{j-1}$. Every term in the later weighted average is at least every term in the earlier one, proving \eqref{eq:iid:block-ratios}.

If raising $I$ decreased $A$, then
\[
\frac ba
<
\frac{A_I(a)}{A_I(b)}
\le
\frac{A_J(a)}{A_J(b)},
\]
and lowering $J$ would increase $A$. Thus a merge that does not decrease $A$ always exists.

Both merges preserve the price order and neither increases the largest fee, so neither increases $D(p_n)$. Choosing a merge that does not decrease $A$ therefore weakly increases actual revenue. Each merge removes a distinct fee level. Repeating the argument leaves a single fee from the original capped vector, with revenue at least as high as before.
\end{proof}

The seller's problem thus reduces from choosing a vector of fees to choosing one common price:
\begin{equation}
\OPT_{\st}
=\max_{0\le p\le c}
p\left(\sum_{j=0}^{n-1}q(p)^j-z(p)^{n-1}\right).
\label{eq:iid:opt}
\end{equation}
Appendix~\ref{app:uniform-proofs} shows that the maximum is attained. The buyer's optimal search responds to revealed values, yet a common fee earns the best static revenue. Identical priors make this possible because uninspected alternatives remain interchangeable. The next section examines how losing this symmetry changes both the complexity of pricing and the revenue obtainable with simple fees.

\section{Static pricing for binary priors}
\label{sec:binary-three}

With heterogeneous priors, uninspected alternatives cease to be interchangeable. Prices shape both which alternatives the buyer investigates and which remain available as blind choices. Binary priors make this tradeoff concrete: each report reveals only a low or a high value, yet coordinating fees is already difficult. The two results in this section show that exact revenue maximization is NP-hard, while comparing three simple pricing schemes secures a constant share of the information surplus. Our computational results use explicitly listed rational endpoints and probabilities, encoded in binary.

\begin{theorem}[Hardness of exact static pricing]
\label{thm:binary:hardness}
For independent nonnegative two-point values, computing a revenue-optimal static fee vector in exact rational representation is NP-hard. Computing optimal static revenue, or deciding whether it reaches a given rational threshold, is also NP-hard. This holds with one deterministic alternative, a group of alternatives sharing their mean and upper value, and one additional two-point alternative with a higher mean and a lower upper value.
\end{theorem}

The reduction starts from Product Partition: divide a list of integers into two groups with equal products. We set each report's low-outcome probability to the reciprocal of an input integer, while keeping these alternatives' means and high values identical. One further alternative has a higher mean but a lower high value. Prices determine which reports the buyer inspects before this alternative and which are left for later. Inspecting it early can end search before a better value is found. Keeping it blind preserves a valuable fallback, but also gives the buyer a strong outside option against which the other reports must be priced. The seller must balance the gains from further search against the surplus needed to induce it.

For each division, revenue is lost through missed improvements in the buyer's choice and surplus left to the buyer. These losses depend on the probabilities that all reports in each group have low outcomes. Independence turns those probabilities into products. Their product is fixed, while the revenue loss increases with their sum, so equal probabilities give the highest revenue. We bound every static fee vector and construct fees attaining this bound for each division. Finding optimal prices therefore solves the partition problem. Appendix~\ref{app:binary-hardness} gives the reduction.

For approximation, the seller can simplify this choice by fixing how a blind fallback will support search. Our algorithm constructs at most three static fee vectors, each with a simple optimal buyer response, and selects the one with the largest revenue guarantee. The construction avoids optimizing over all the search patterns that different fees could induce.

\begin{theorem}[A static $3$-approximation]
\label{thm:binary-three}
For independent nonnegative two-point values, a deterministic static fee vector earning at least $\Delta/3$ can be computed in polynomial time from explicit rational endpoints and probabilities. This is a $3$-approximation to both optimal static revenue and the information-surplus bound.
\end{theorem}

The seller captures at least a third of the gain from learning, regardless of the number of alternatives or the degree of heterogeneity. Each proposed fee vector induces a simple optimal search policy, so its revenue guarantee can be calculated directly from the priors. We build on the surplus decomposition and reserve construction of~\cite{chawla2026}, then use two further schemes to capture the remaining information value.

\subsection{Surplus and the reserve scheme}

The schemes follow two ways information can improve the buyer's choice. Imagine first revealing all values except that of the best blind alternative. Its mean remains a reserve, and the other alternatives improve the choice when they exceed it. Revealing the reserved value then allows the buyer to check whether this fallback is actually preferable. Separating these two gains suggests how to price the reports.

Each alternative has a high value $h_i$, occurring with probability $\theta_i$, and a low value $l_i$, occurring otherwise. Its realized value is $X_i$, where $0\le l_i\le h_i$ and $0\le\theta_i\le1$.
Order alternatives by their prior means, $\mu_1\ge\mu_2\ge\cdots\ge\mu_n$, so alternative $1$ is a best blind choice. Denote the two highest prior means by $m=\mu_1=\mu_*$ and $t=\mu_2$, and alternative $1$'s low value and its probability by $s=l_1$ and $a=1-\theta_1$. The best competing value is $Y=\max_{i\ge2}X_i$; after excluding alternative $2$ as well, it is $Z=\max_{i\ge3}X_i$. For $n=2$, the latter maximum is $Z=0$. When report $i$ is the only remaining source of information, its information value at a nonnegative fallback $y$ is
\begin{equation}
v_i(y)=\E[\max\{X_i,y\}]-\max\{\mu_i,y\}.
\label{eq:binary:voi}
\end{equation}
This thought experiment separates the surplus into the additional value $A$ of report $1$ and the gain $B$ from searching while it remains blind:
\begin{equation}
\Delta=A+B,\qquad
A=\E[v_1(Y)],\qquad B=\E[(Y-m)^+].
\label{eq:binary-AB}
\end{equation}
The reserve scheme collects the search gain $B$ while keeping the best blind choice intact. The other schemes address $A$, the value of inspecting that alternative itself. A large prior mean alone does not tell the seller which gain dominates: the other alternatives may offer substantial upside, or most of the useful information may concern the best blind choice. The algorithm therefore keeps both sources of revenue in view.

\begin{table}[t]
\centering
\caption{Three static pricing schemes when the second-highest prior mean exceeds alternative $1$'s low value ($t>s$). Each row gives a scheme; the three middle columns specify its fixed report fees, and the last gives an expected-revenue lower bound. Alternatives $1$ and $2$ have the highest prior means, $m$ and $t$. Alternative $1$ has low value $s$ with probability $a$; alternative $2$ has high value $h_2$ with probability $\theta_2$. The best realized values among alternatives $2$ onward and $3$ onward are $Y$ and $Z$, respectively, and $v_1(y)$ is report $1$'s information value against a known competing value $y$. An infinite fee withholds a report while preserving blind selection.}
\label{tab:binary}
\label{tab:binary-intuition}
\small
\setlength{\tabcolsep}{5pt}
\renewcommand{\arraystretch}{1.12}
\begin{tabular*}{\textwidth}{@{\extracolsep{\fill}}lcccl@{}}
\toprule
& \multicolumn{3}{c}{\textbf{Fixed report fee}} & \\
\cmidrule(lr){2-4}
\textbf{Scheme} & \textbf{Report $1$} & \textbf{Report $2$} & \textbf{Report $i\ge3$} & \textbf{Revenue guarantee} \\
\midrule
Reserve scheme & $\infty$ & $\E[(X_2-m)^+]$ & $\E[(X_i-m)^+]$ & $B=\E[(Y-m)^+]$ \\
\addlinespace[4pt]
Early-report scheme & $a(t-s)$ & $\infty$ & $\E[(X_i-t)^+]$ & $C=a(t-s+\E[(Z-t)^+])$ \\
\addlinespace[4pt]
Free-report scheme & $v_1(h_2)$ & $0$ & $\infty$ & $D=\theta_2v_1(h_2)$ \\
\bottomrule
\end{tabular*}
\end{table}

The reserve scheme in Table~\ref{tab:binary} withholds report $1$ and prices every other report at its expected improvement over the blind payoff $m$. No other blind choice improves on this reserve, and charging each report's full expected gain prevents any search policy from achieving utility above $m$. An optimal response inspects in decreasing order of high values. Low outcomes leave the fallback unchanged; at the first outcome above $m$, the buyer can stop because no remaining report for sale can reveal a higher value. This search retains utility $m$ while realizing the full gain $B$, which becomes revenue. The buyer is indifferent between searching and taking the blind choice, so our seller-favorable tie-breaking supports the guarantee. The next lemma establishes optimality and also supplies the continuation prices for the second scheme.

\begin{lemma}[Binary reports above a common reserve]
\label{lem:binary:common}
Suppose a permanent fallback has value $b$ and the remaining independent two-point alternatives satisfy $\mu_i\le b$. At prices $p_i=\E[(X_i-b)^+]$, buyer utility is $b$. There is an optimal response that inspects in decreasing order of $h_i$, stops at the first value strictly above $b$, and otherwise selects the fallback. Its revenue is
$\E\!\left[(\max_iX_i-b)^+\right].$
The statement also applies when these alternatives can be selected blindly.
\end{lemma}
\begin{proof}
Blind selection yields at most $b$, so it is redundant. Applying Lemma~\ref{lem:binary:capped}, which bounds utility by capped values, with every $g_i=b$ gives an upper bound of $b$; stopping attains it.

Ignore alternatives with $h_i\le b$, whose fees are zero and whose values cannot improve the fallback. For each remaining alternative, $l_i\le\mu_i\le b<h_i$. On a lower outcome the proposed policy continues with utility $b$; on an upper outcome it stops. Backward induction gives net utility
\[
-p_i+\theta_i h_i+(1-\theta_i)b
=-\theta_i(h_i-b)+\theta_i h_i+(1-\theta_i)b=b
\]
at each step. Decreasing order of $h_i$ ensures that the first upper outcome is at least every possible uninspected value. The selected value therefore equals $\max\{b,\max_iX_i\}$ in this auxiliary problem. Subtracting utility $b$ gives the stated revenue.
\end{proof}

\subsection{Two further pricing schemes}

Selling report $1$ means giving up the best blind choice. The early-report scheme supports this purchase by preserving alternative $2$ as a fallback. In the main case $t>s$, it keeps this alternative blind at its mean $t$ and charges $p_1=a(t-s)$ for report $1$. The reserve protects the buyer against a disappointing report: after a low outcome, the buyer can choose the reserve $t$ instead of value $s$. This improvement occurs with probability $a$, so the fee collects its expected value and leaves utility
\begin{equation}
\E\max\{X_1,t\}-p_1
=\theta_1h_1+a t-a(t-s)=m.
\label{eq:binary:blind-replacement}
\end{equation}
Thus the buyer can purchase report $1$ without giving up the original blind payoff $m$. Price the remaining reports at their expected improvements over $t$, so the same reserve also supports later sales.

An optimal response inspects report $1$ first and stops on a high outcome. The revealed value is then at least $t$; since the remaining fees already cover improvements over $t$, further search cannot raise net utility. After a low outcome, the buyer searches the remaining reports in decreasing order of high values, retaining alternative $2$ as the fallback. Lemma~\ref{lem:binary:common} gives continuation utility $t$ and revenue $\E[(Z-t)^+]$. The guarantee ${C=a(t-s)+a\E[(Z-t)^+]}$ combines the initial payment with this continuation revenue, weighted by the low-outcome probability. The whole response leaves utility $m$. Appendix~\ref{app:binary-proofs} shows that no other search policy gives higher utility, including policies that inspect other reports first.

The free-report scheme uses free information to encourage a paid purchase. It makes report $2$ free, charges the fixed fee $v_1(h_2)$ for report $1$, and withholds all other reports. Taking the free information first is optimal because it can be ignored. With only one paid report available, the buyer then compares its fee with its expected improvement in choice value. When the free report reveals $h_2$, this becomes the best competing choice, making report $1$ worth exactly its fee. Buying it is therefore optimal, giving the revenue guarantee $D=\theta_2v_1(h_2)$. Free disclosure earns revenue indirectly by making this paid comparison possible; purchases after the low outcome can only add to the guarantee.

Table~\ref{tab:binary} collects the three schemes for $t>s$. If $t\le s$, even alternative $1$'s low value is a sufficient fallback for all other blind choices. The seller can reveal report $1$ for free and price the others against that low value. Together with the reserve scheme, this gives the two candidates used in this case.

Consider three independent alternatives with equally likely endpoint pairs $(2,10)$, $(1,9)$, and $(0,8)$. Their means are $6$, $5$, and $4$, giving $m=6$, $t=5$, $s=2$, and $a=1/2$. The reserve scheme posts $(\infty,1.5,1)$, the early-report scheme $(1.5,\infty,1.5)$, and the free-report scheme $(0.5,0,\infty)$. If the free report reveals $9$, inspecting alternative $1$ improves expected choice value by $0.5$.

The revenue guarantees are $B=2$, $C=2.25$, and $D=0.25$, so the algorithm selects the early-report scheme. The buyer pays $1.5$ for report $1$ and stops if it reveals $10$. After a low outcome of $2$, alternative $2$ remains a blind reserve worth $5$, and the buyer purchases report $3$ for $1.5$. Expected revenue is ${1.5+(1/2)1.5=2.25}$. Full information gives expected value $8.5$, so the surplus is $\Delta=2.5$ and the scheme captures $90\%$.

Here report $1$ tests whether the best blind alternative is already good enough; report $3$ becomes useful when that test disappoints. Keeping alternative $2$ blind supports both purchases.

\subsection{The approximation guarantee}

The three schemes are designed to perform well in different instances. When most of the surplus comes from searching around the best blind choice, the reserve scheme captures that gain. When this source is small, most of the surplus comes from inspecting the reserved alternative itself. The other two schemes cover that value together, through an early sale or a paid comparison after free disclosure.

To establish this coverage, consider how a binary report's value changes with the competing option. Information is most useful when that option equals the report's prior mean, where the choice is hardest to make without inspection. For report $i$, information has no value below its low outcome $l_i$. Its value then rises at rate $1-\theta_i$ up to the mean and falls as the competing option becomes more attractive. Two bounds capture this shape:
\begin{align}
v_i(y)&\le(1-\theta_i)(y-l_i)^+,
\label{eq:binary-tent-upper}\\
v_i(y)-v_i(x)&\le(1-\theta_i)(y-x)\qquad (y\ge x).
\label{eq:binary-tent-lipschitz}
\end{align}
The first bounds the report's information value; the second bounds how much a better competing option can increase willingness to pay. Both depend on the probability of a low outcome, when a better fallback matters for the buyer's choice. Appendix~\ref{app:binary-proofs} gives the full piecewise formula.

Apply these bounds to report $1$, conditioning on alternative $2$'s outcome. On a low outcome, its information value is bounded by the protection from reserve $t$ and the gains available above that reserve, the two terms in the early-report guarantee $C$. On a high outcome, the free-report fee covers the comparison with $h_2$; the slope bound controls any extra value from better competing alternatives using the same continuation term in $C$. Averaging gives $A\le C+D$. Adding the reserve revenue $B$ covers the full surplus, so choosing the largest of these three guarantees captures at least a third. The seller computes these bounds before search and posts a single selected fee vector for the entire interaction.

\begin{proof}[Proof of Theorem~\ref{thm:binary-three}]
If $t\le s$, the reserve vector yields at least $B$ and the free-report-$1$ vector yields at least $A$. Choosing the larger of their certified bounds gives $\Delta/2$, which is stronger than required.

Suppose $t>s$. On the lower Bernoulli branch of alternative $2$, $l_2\le t$ and~\eqref{eq:binary-tent-upper} give
\[
v_1(\max\{l_2,Z\})\le a\bigl(t-s+(Z-t)^+\bigr).
\]
On the upper branch, $h_2\ge t$ and the upward-slope bound~\eqref{eq:binary-tent-lipschitz} give
\[
v_1(\max\{h_2,Z\})
\le v_1(h_2)+a(Z-h_2)^+
\le v_1(h_2)+a(Z-t)^+.
\]
The mixture calculation remains valid if the two endpoints coincide. Independence of $X_2$ and $Z$ now implies
\[
A\le(1-\theta_2)a(t-s)+a\E[(Z-t)^+]+\theta_2v_1(h_2)\le C+D.
\]
Using~\eqref{eq:binary-AB}, we obtain
${\Delta=A+B\le B+C+D}$.
One certified revenue is therefore at least $\Delta/3$.

For the algorithm, sort the means, calculate the appropriate two or three lower bounds, and post the vector with the largest lower bound. Expectations of maxima of independent two-point values are computable by sorting their endpoints and multiplying cumulative probabilities on the resulting intervals. This computes $A,B,C,D$ and every fee exactly in polynomial time for rational input. An unavailable report can equivalently be assigned the finite rational fee $1+\max_i h_i$, which is above every individual information cap and is never bought. The information-surplus bound of Lemma~\ref{lem:model:surplus} gives $\OPT_{\st}\le\Delta$, completing the approximation claim.
\end{proof}

Whether static prices can always capture more than one third of the surplus remains open. A matching lower bound would have to limit every static fee vector, beyond the three schemes considered here. Exact optimization is NP-hard, while the best polynomial-time approximation to optimal static revenue remains open.

\section{The static--adaptive revenue gap}
\label{gen:section}

The binary schemes capture a constant share of the surplus with static fees. To extend the analysis to general priors, we compare static and optimal adaptive revenue. An adaptive rule allows fees to change with the public inspection history. The seller announces it before search, so the buyer evaluates purchases together with future fees. Denote optimal adaptive revenue by $\OPT_{\dyn}$ and the worst-case revenue gap by $G_n=\sup\OPT_{\dyn}/\OPT_{\st}$, where the supremum ranges over $n$-alternative instances with positive surplus.

\begin{theorem}[Sharp asymptotic adaptivity gap]
\label{gen:adaptivegap}
For independent nonnegative finite-mean values and every $n\ge2$, uniform static pricing guarantees
\[
\OPT_{\st}\ge\OPT_{\unif}\ge\frac{\OPT_{\dyn}}{1+\ln n}.
\]
Moreover, the worst-case adaptivity gap satisfies
\[
\frac{H_n-1}{1-(1-e^{-1})^{n-1}}\le G_n\le1+\ln n,
\]
where $H_n=\sum_{k=1}^n1/k$. In particular, $G_n/\ln n\to1$ as $n\to\infty$.
\end{theorem}

We bound uniform-price revenue through the information surplus, show that adaptive pricing extracts this surplus, and give an exponential instance attaining the logarithmic gap against every static fee vector.

\begin{figure}[t]
\centering
\includegraphics[width=\textwidth]{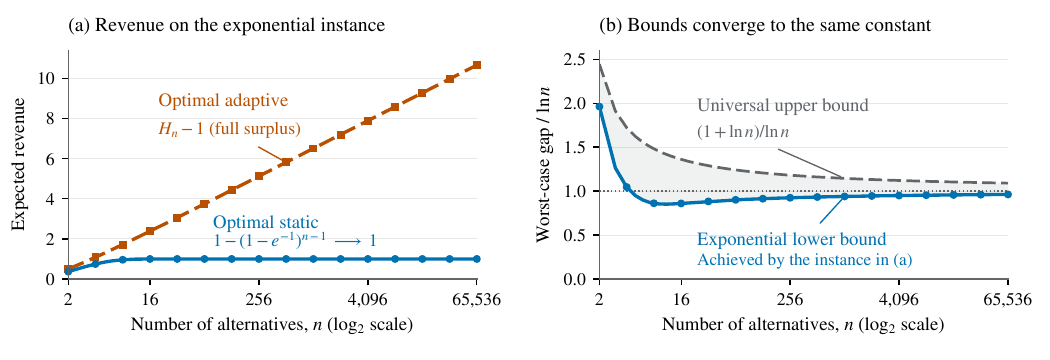}
\caption{The logarithmic revenue gap. (a) With mean-one exponential priors, optimal static revenue approaches $1$, while adaptive revenue grows as $H_n-1$. (b) The lower and upper bounds on $G_n/\ln n$ both approach $1$, showing asymptotic tightness. Shading marks the region between the bounds in Theorem~\ref{gen:adaptivegap}; horizontal axes use a base-two logarithmic scale.}
\Description{Two panels for n from 2 to 65536 on logarithmic horizontal axes. Left: optimal static revenue approaches one while optimal adaptive revenue grows to about 10.67. Right: the exponential instance gives a lower bound on the worst-case gap divided by log n; the universal guarantee gives an upper bound. Both converge to one. Shading marks the region between these deterministic bounds, not statistical uncertainty.}
\label{fig:gap}
\end{figure}

\subsection{A uniform-price revenue bound}

Let demand $Q(t)$ be the expected number of purchases at a common fee $t$. For a fixed search policy, expected utility is linear in the fee, with slope equal to minus the expected purchase count. The buyer chooses the policy giving the highest utility, so optimal utility traces the upper envelope of these lines. Its negative slope therefore records demand even as the preferred policy changes. When reports are free, the buyer can learn every value; at a sufficiently high fee, choosing blindly is optimal. Integrating demand recovers this utility difference, $\Delta=\int_0^\infty Q(t)\,dt$. Revenue at one fee is the rectangle $tQ(t)$ under the demand curve, so the best uniform revenue ${R_*=\OPT_{\unif}}$ is the largest such rectangle.

Two restrictions control the area: the buyer can purchase at most $n$ reports, and any fee earns at most $R_*$. Thus $Q(t)\le\min\{n,R_*/t\}$. Before any value is observed, the buyer either declines all reports or purchases at least one. A fee above $R_*$ therefore cannot attract demand. Integrating gives
\[
\Delta\le\int_0^{R_*/n}n\,dt+\int_{R_*/n}^{R_*}\frac{R_*}{t}\,dt=R_*(1+\ln n).
\]
Below $R_*/n$, the cap of $n$ purchases bounds the area by $R_*$. The logarithm comes from the remaining price range: between $R_*/n$ and $R_*$, each doubling of the fee contributes at most the same amount, and the number of these intervals grows logarithmically with $n$. The argument applies across general priors without characterizing every search policy; Appendix~\ref{app:demand-integral} gives the formal derivation.

\subsection{Full extraction with a single initial payment}
\label{sec:dynamic}

Adaptive pricing attains the information-surplus benchmark through a simple access fee: charge $\Delta$ for the first report and make every subsequent report free. If the purchase count at history $h$ is $k(h)$, the fee for each remaining report $i$ is
\begin{equation}
 p_i(h)=\Delta\,\mathbf{1}\{k(h)=0\}.
 \label{gen:dynamicfee}
\end{equation}

\begin{proposition}[Full extraction under public commitment]
\label{gen:dynamicexact}
For independent nonnegative finite-mean values, the committed fee rule~\eqref{gen:dynamicfee} earns revenue $\Delta$ under seller-favorable tie-breaking. Buying any first report, inspecting all remaining reports for free, and selecting the highest revealed value is an optimal buyer policy. The buyer's initial expected utility is $\mu_*$, and $\OPT_{\dyn}=\Delta$.
\end{proposition}

The buyer knows that the first payment unlocks all remaining reports at no further cost. Revealing every value is then an optimal continuation, giving initial net utility ${\E[\max_iX_i]-\Delta=\mu_*}$. Paying for access leaves the buyer the same payoff as choosing blindly, while the seller collects the entire improvement in choice value. A small discount on the initial fee makes purchase strictly preferable when the surplus is positive and brings revenue arbitrarily close to full extraction (Appendix~\ref{app:dynamic-proof}).

This is a two-part tariff: the initial payment collects willingness to pay for access, and a zero marginal price encourages efficient information acquisition. Since reports have no production cost, the marginal price equals marginal cost. The first purchase includes the value of future learning, which explains why its fee can exceed the static single-report cap in Lemma~\ref{gen:cap}. The rule needs to record only whether a purchase has occurred. It already reaches the revenue upper bound, so conditioning later fees on the revealed values cannot improve revenue.

\subsection{A matching exponential instance}

An instance with identical priors already attains the logarithmic gap. Take independent mean-one exponential values. The blind payoff is one, while the expected maximum is $H_n$, giving adaptive revenue $H_n-1$. By Theorem~\ref{thm:iid:uniform}, a common fee is optimal among all static fee vectors. The best fee is $e^{-1}$ and earns $1-(1-e^{-1})^{n-1}$ (Appendix~\ref{app:commitment-proofs}). At this fee, a draw above one ends paid search, and each inspection produces such a draw with probability $e^{-1}$. The expected number of purchases therefore stays bounded as more alternatives are added.

Figure~\ref{fig:gap}(a) shows the resulting separation: optimal static revenue approaches one, while adaptive revenue grows logarithmically with the number of alternatives. Panel~(b) divides the exponential lower bound and the universal upper bound on the revenue gap by $\ln n$. Both converge to one, identifying the leading constant in the logarithmic loss and completing Theorem~\ref{gen:adaptivegap}. The timing of payment explains the gap: per-report fees lose future sales when search ends, whereas an access fee collects the value of learning before search begins.

\section{Discussion and conclusion}
\label{sec:conclusion}

Our results show how the option to choose without inspection shapes information pricing. A common fee maximizes static revenue for identical priors. Heterogeneity makes exact optimization NP-hard even for binary priors, yet selecting among three simple schemes gives a $3$-approximation to the information surplus. For general priors, uniform pricing gives a tight logarithmic approximation to optimal adaptive revenue. A committed access fee followed by free reports extracts the full surplus. These guarantees provide a basis for monetizing agent-mediated search through information sales and quantify the revenue cost of keeping report fees fixed.

For binary priors, it remains open whether static prices can always capture more than one third of the surplus. The best polynomial-time approximation to optimal static revenue also remains open; for general heterogeneous finite-support priors, even a constant-factor approximation is unknown. Extensions to correlated values, partial disclosure, and competing information providers would clarify the scope of these guarantees.

\clearpage
\bibliographystyle{ACM-Reference-Format}
\bibliography{references}

\clearpage
\appendix
\section{Model details and basic bounds}
\label{app:model-details}

Write $[n]=\{1,\ldots,n\}$, $x^+=\max\{x,0\}$, and $M=\E[\max_iX_i]$, so $\Delta=M-\mu_*$. These expectations are finite because $\max_iX_i\le\sum_iX_i$. For identical priors, write $\OPT_n=\OPT_{\st}$ and $\mathcal R_n(p)=R(p\mathbf1)$.
Finite-horizon backward induction implements seller-favorable ties by maximizing continuation utility first and expected payments second; a deterministic policy suffices. Free reports can be inspected first because their information can be ignored.

\subsection{The information-surplus bound}
\begin{lemma}[Information-surplus bound]
\label{lem:model:surplus}
Any mechanism with voluntary payments in which the buyer can stop initially for $\mu_*$ has expected revenue at most $\Delta$. Uniform and arbitrary static revenues therefore satisfy
${0\le\OPT_{\unif}\le\OPT_{\st}\le\Delta}$.
\end{lemma}
\begin{proof}
Expected selected value is at most $M$, and voluntary participation gives utility at least $\mu_*$. Revenue is their difference.
\end{proof}

\subsection{Individual information caps}
\label{app:cap-proof}
\begin{proof}[Proof of Lemma~\ref{gen:cap}]
Fix a history with report $i$ uninspected and best observed value $b$. Let $f(x)$ be optimal continuation utility using the other alternatives and a permanent fallback $\max\{b,x\}$. Each fixed policy's payoff is nondecreasing and $1$-Lipschitz in $x$, since only one alternative is selected. Taking the optimum preserves these properties, so $f(x)-f(\mu_i)\le(x-\mu_i)^+$. Inspection therefore yields at most $f(\mu_i)+c_i-p_i$, while keeping $i$ blind forever gives $f(\mu_i)$. The latter strictly dominates when $p_i>c_i$ and weakly dominates at equality.

When lowering a fee above $c_i$ to the cap, replace each eventual inspection by this no-inspection continuation. Backward induction shows that optimal utility cannot rise. The original optimal policy never bought $i$ and remains feasible with the same utility and payments, so seller-favorable ties weakly increase revenue. Apply this argument one report at a time.
\end{proof}

\section{Deferred details for uniform optimality}
\label{app:uniform-proofs}

Use the quantities defined in Section~\ref{sec:iid}. If the prior is
constant, all revenues are zero. Otherwise $c>0$, $\mu>0$, and
$0<F(\mu-)\le F(\mu)<1$, where $F(x-)=\Prb[X<x]$.

\subsection{Proof of the exact revenue formula}
\label{app:revenue-proof}
With best inspected value $b\ge0$, the last report's value is ${w(b)=\E[\max\{b,X\}]-\max\{b,\mu\}}$, or equivalently
\begin{equation}
w(b)=
\begin{cases}
\mathbb E[(b-X)^+],&b\le\mu,\\
\mathbb E[(X-b)^+],&b\ge\mu.
\end{cases}
\label{eq:iid:last-value}
\end{equation}
The maximum of $w$ is $w(\mu)=c$.

\begin{proof}[Proof of Proposition~\ref{prop:iid:revenue}]
First suppose all fees are positive. Use the cheapest-first response from Lemma~\ref{lem:iid:order}, resolving equal prices by a fixed label order. Suppose at least two alternatives remain, the cheapest remaining price is $p$, and the best inspected value is $b$.

If $b\le\mu$, inspecting the cheapest alternative and then stopping while retaining one uninspected alternative gives utility
$\mathbb E[\max\{X,\mu\}]-p =\mu+c-p\ge\mu.$
Thus inspection is weakly worthwhile. At equality, a positive price gives positive revenue and is favored over stopping.

If $b\ge\mu$, inspection followed by stopping gives utility improvement
$\mathbb E[(X-b)^+]-p$. This is nonnegative exactly when $b\le g(p)$. At equality the inspection is utility-neutral and can generate revenue.

If $b>g(p)$, the observed value dominates every blind choice. For a policy inspecting remaining alternative $i$ with probability $\beta_i$, expected improvement over $b$ is at most $\sum_i\beta_i\E[(X-b)^+]$: the realized gain is bounded by the sum of inspected positive excesses, and inspection is independent of the unrevealed value. Since ${\E[(X-b)^+]<p\le p_i}$, any policy that inspects with positive probability pays more than this gain. Stopping is therefore strictly optimal.

It follows that, among the first $n-1$ alternatives, alternative $j$ is inspected exactly when all earlier realized values are at most $g(p_j)$. Because $g(p_1)\ge\cdots\ge g(p_n)$, this event automatically implies all earlier continuation decisions. Its probability is
$q(p_j)^{j-1}$.

For the last report, equation~\eqref{eq:iid:last-value} makes inspection optimal exactly for ${\ell(p_n)\le b\le g(p_n)}$, including the endpoints by seller-favorable ties. This event guarantees that all earlier alternatives were reached and has probability $q(p_n)^{n-1}-z(p_n)^{n-1}$. Summing payments proves \eqref{eq:iid:revenue-vector}.

Inspect zero-priced reports first; their information can be ignored, so this does not restrict continuation policies. They contribute no revenue, and the same event formula applies to later positive-price reports. If $p_n=0$, both sides of~\eqref{eq:iid:revenue-vector} are zero.
\end{proof}

\subsection{Attainment of the optimal uniform price}
\begin{proof}[Proof of attainment in Theorem~\ref{thm:iid:uniform}]
On $(0,c]$, the thresholds $g(p)$ and $\ell(p)$ are continuous. The right-continuity of $F$ implies that $q(p)=F(g(p))$ is upper semicontinuous. The left-continuity of $F(x-)$ implies that $z(p)=F(\ell(p)-)$ is lower semicontinuous. Thus the expression in \eqref{eq:iid:uniform-revenue} is upper semicontinuous on $(0,c]$. At $p=0$, its limit is zero because revenue is bounded by $np$. It therefore attains its maximum on the compact interval $[0,c]$.
\end{proof}

\section{Proofs for binary priors}
\label{app:binary-proofs}

We use the notation of Section~\ref{sec:binary-three}.

\subsection{Hardness of exact static pricing}
\label{app:binary-hardness}

\begin{proof}[Proof of Theorem~\ref{thm:binary:hardness}]
We reduce from the following product-partition problem: given integers $b_1,\ldots,b_m\ge2$ with product $K^2$, find a subset with product $K$. For completeness, this problem is NP-hard by Exact Cover by 3-Sets~\cite[problem SP2]{garey1979}. Assign distinct primes to universe elements, let each triple's number $x_j$ be the product of its primes, and let $T$ be the product of all universe primes. Exact covers correspond to subsets with product $T$. Writing $X=\prod_jx_j$, append $X$ and $T^2$ to the list. Its product is $(XT)^2$. A subset with product $XT$ must contain exactly one appended number: with $X$, its original numbers multiply to $T$; with $T^2$, their complement does. The converse is immediate. Nontrivial instances have all numbers at least two, and this construction has polynomial bit length.

Set $Q=K^{-2}$, $d=(4\max_i b_i)^{-1}$, and $a=dQ/16$. Include a deterministic value $1$ and an auxiliary alternative $A$ with equiprobable values $1+2a-d$ and $h=1+d$, hence mean $\mu=1+a$. Each other alternative $i\in[m]$ has high value $H=1+d+2a$ with probability $q_i=1-1/b_i$, and low value $l_i=1-(b_i-1)(d+2a)$, hence mean $1$. All low values lie in $(0,1)$, since $a\le d/16$ and $(b_i-1)(d+2a)<9/32$. All parameters are rational with polynomial bit length. Write $Q_S=\prod_{i\in S}(1-q_i)$, with $Q_\varnothing=1$. Full-information welfare is $W=H-(d/2+2a)Q$, and surplus is $\Delta=W-\mu$.

Fix arbitrary nonnegative fees. Define reservation caps $c_i=\max\{1,H-p_i/q_i\}$ and $z=\max\{1,h-2p_A\}$; unavailable reports have cap $1$. With independent high-outcome indicators $B_i,B_A$, the buyer's utility is $\max\{U_0,U_1\}$, where
\begin{equation}
\begin{aligned}
U_0&=\E\max\{\mu,c_iB_i:i\in[m]\},\\
U_1&=\E\max\{1,zB_A,c_iB_i:i\in[m]\}.
\end{aligned}
\label{eq:hardness:utilities}
\end{equation}
Here $U_0$ keeps $A$ blind, while $U_1$ is mandatory search with fallback $1$; both formulas follow from the reservation-index rule~\cite{weitzman1979}. Clamping inactive indices to $1$ leaves these utilities unchanged. To see why the two cases exhaust search with nonobligatory inspection, a deterministic policy has a single all-low path before the first high outcome. A high $H$ ends useful search. If the policy inspects $A$ on this path, it never needs $A$ as a blind choice; otherwise it keeps $A$ blind. Other blind means are dominated by the permanent fallback. Zero-fee inspections after a terminal high can be deleted. Randomization cannot improve either utility or revenue among utility optima. If $z\le\mu$, the maximum defining $U_1$ is pointwise below that for $U_0$, strictly so when all outcomes are low. Thus the mandatory branch can be optimal only if $z>\mu$.

Seller-favorable ties require care. For any deterministic policy $f$, discounting all fees gives utility $U_f((1-\varepsilon)p)=U_f(p)+\varepsilon R_f(p)$. There are finitely many policies, so a constant-policy limit of discounted optima maximizes revenue among the original utility optima. Within mandatory search, choose reservation-index policies for these discounted problems. At a tied active cap $c$, the discounted caps are $c+\varepsilon(H-c)$ for report $i$ and $c+\varepsilon(h-c)$ for $A$. Since $H>h$, reports with $c_i\ge z$ precede $A$ in such a limit. Reports with smaller caps follow it only after its low outcome; inactive reports may be omitted. Including them can only increase the gross-welfare bound below.

A policy keeping $A$ blind has welfare at most $H(1-Q)+\mu Q$ and utility at least $\mu$, hence revenue at most $R_0=\Delta-(d/2-a)Q$. Otherwise let $S=\{i:c_i\ge z\}$, $T=[m]\setminus S$, $u=Q_S$, and $v=Q_T$, so $uv=Q$. Put $M=\max\{1,c_iB_i:i\in T\}$, $e=\E M$, and $\phi=\E(\mu-M)^+$. As $M\le z$, conditioning on all reports in $S$ being low gives $U_1-U_0=u[(z-e)/2-\phi]$. Optimality of this branch implies $z\ge e+2\phi$, and therefore
\begin{equation}
U_1\ge(1-u/2)z+(u/2)e\ge e+(2-u)\phi
       \ge\mu+av(1-u).
\label{eq:hardness:rent}
\end{equation}
The last inequality uses $x+(2-u)(\mu-x)^+\ge\mu$ for $x\ge1$, with value $\mu+a(1-u)$ at $x=1$; all reports in $T$ are low with probability $v$. Inspecting all of $S$, then $A$, then all of $T$ after an auxiliary low has welfare $W_S=W-a(u-Q)$. Its only loss from full information is $H-h=2a$ when $S$ is low, $A$ is high, and some report in $T$ is high. Omitting inactive reports cannot improve welfare, so the revenue-favorable index policy has welfare at most $W_S$. Subtracting~\eqref{eq:hardness:rent} bounds revenue by $\Delta-a(u+v-2Q)$ for every fee vector using this branch.

Each subset attains this bound. Choose the auxiliary cap $z^*=1+a(1+v)$, set $c_i=z^*$ on $S$ and $c_i=\mu$ on $T$, and charge $p_i=q_i(H-c_i)$ and $p_A=(h-z^*)/2$. These fees are nonnegative. Then $e=1+a(1-v)$ and $\phi=av$, so $U_0=U_1=\mu+av(1-u)$. Searching $S$, then $A$, then $T$ after an auxiliary low is utility-optimal, has welfare $W_S$, and earns $R(S)=\Delta-a(u+v-2Q)$. Seller-favorable ties give at least this revenue. Moreover, $\Delta-R_0=8a-aQ\ge7a$, whereas $\Delta-R(\varnothing)=a(1-Q)\le a$. Thus the blind branch is dominated, and
\begin{equation}
\OPT_{\st}=\Delta-a\left(\min_{S\subseteq[m]}\{Q_S+Q_{[m]\setminus S}\}-2Q\right).
\label{eq:hardness:optimum}
\end{equation}

Let $r^*=\Delta-2a(K^{-1}-Q)$. Since $Q_SQ_T=K^{-2}$, their sum is at least $2/K$, with equality exactly when $\prod_{i\in S}b_i=K$. Equation~\eqref{eq:hardness:optimum} therefore gives $\OPT_{\st}\ge r^*$ exactly on yes instances, proving value and threshold hardness with polynomial-size rational inputs.

Finally, an exact optimal fee vector also solves the source problem without a revenue oracle. Compute its caps; reject if $z\le\mu$, and otherwise test whether the subset $S=\{i:c_i\ge z\}$ has product $K$. On a yes instance, optimal revenue is $r^*>R_0$, so the mandatory branch is optimal. Its all-price upper bound forces ${r^*\le\Delta-a(Q_S+Q_T-2Q)\le r^*}$, hence this subset is balanced. No subset can pass on a no instance. This extraction uses only exact rational comparisons and integer products.
\end{proof}

\subsection{The shape of a binary report's value}
Direct calculation gives
\begin{equation}
v_i(y)=
\begin{cases}
(1-\theta_i)(y-l_i),&l_i\le y\le\mu_i,\\
\theta_i(h_i-y),&\mu_i\le y\le h_i,\\
0,&\text{otherwise}.
\end{cases}
\label{eq:binary:tent}
\end{equation}
The formula also covers coincident endpoints. Extending its rising segment gives~\eqref{eq:binary-tent-upper}; every segment has slope at most $1-\theta_i$, giving~\eqref{eq:binary-tent-lipschitz}.

\subsection{A capped-value bound for a permanent reserve}

The following bound applies when all blind choices except one reserve are redundant.

\begin{lemma}[A capped-value bound]
\label{lem:binary:capped}
Consider independent alternatives indexed by a set $T$ and a permanent deterministic fallback $b\ge0$. Suppose only inspected alternatives can be selected in place of that fallback. For numbers $g_i\ge0$, set
$p_i=\E[(X_i-g_i)^+]$. Every search policy has utility at most
\[
\E\!\left[\max\{b,\max_{i\in T}\min\{X_i,g_i\}\}\right].
\]
\end{lemma}
\begin{proof}
Let $I\subseteq T$ be the random set of inspected alternatives. Since $i\in I$ is determined before observing $X_i$, independence gives $\E[\sum_{i\in I}p_i]=\E[\sum_{i\in I}(X_i-g_i)^+]$. Subtracting these excesses from $\max\{b,\max_{i\in I}X_i\}$ removes at least the maximizing alternative's excess, leaving at most $\max\{b,\max_{i\in T}\min\{X_i,g_i\}\}$. Taking expectations proves the claim.
\end{proof}

\subsection{Three static constructions}

We verify the guarantees for Table~\ref{tab:binary}, including the alternative construction for $t\le s$.

\begin{lemma}[Keeping the highest-mean alternative blind]
\label{lem:binary-reserve}
Set $p_1=\infty$ and $p_i=\E[(X_i-m)^+]$ for $i\ge2$. Revenue is at least $B$.
\end{lemma}
\begin{proof}
Alternative $1$ supplies a permanent blind reserve $m$, dominating every other blind mean. Lemma~\ref{lem:binary:common} gives an optimal response with utility $m$ and selected value $\max\{m,Y\}$, hence revenue $B$. Seller-favorable ties yield at least this amount.
\end{proof}

\begin{lemma}[Charging for the highest-mean report first]
\label{lem:binary-primary}
If $t\le s$, a static vector obtains revenue at least $A$. If $t>s$, the second vector in Table~\ref{tab:binary} obtains revenue at least $C=a\bigl(t-s+\E[(Z-t)^+]\bigr)$.
\end{lemma}
\begin{proof}
First suppose $t\le s$. Make report $1$ free and set $p_i=\E[(X_i-s)^+]$ for $i\ge2$. Inspecting the free report first is optimal. On $X_1=s$, an event of probability at least $a$, the fallback $s$ dominates all remaining blind means, so Lemma~\ref{lem:binary:common} gives conditional revenue ${\E[(Y-s)^+]}$. Other payments are nonnegative, giving revenue at least
$a\E[(Y-s)^+]\ge\E[v_1(Y)]=A,$
where the inequality is the pointwise bound~\eqref{eq:binary-tent-upper}.

For $t>s$, use the second vector in Table~\ref{tab:binary}. The blind reserve $t$ dominates every other blind mean except $m$. Whenever alternative $1$ remains uninspected, inspecting it and choosing between its value and the reserve yields utility $m$ by~\eqref{eq:binary:blind-replacement}, even if all other observations are ignored. Thus any terminal blind choice of alternative $1$ can be replaced by inspection. Optimal utility is unchanged if alternative $2$ becomes a permanent fallback $t$ and all other alternatives require inspection before selection.

Since $m\ge t>s$ implies $\theta_1>0$, define $g_1=h_1-a(t-s)/\theta_1\ge t$ and $g_i=t$ for $i\ge3$. These caps satisfy $p_i=\E[(X_i-g_i)^+]$. Every other capped value is at most $t$, so Lemma~\ref{lem:binary:capped} bounds utility by $\E\max\{t,\min\{X_1,g_1\}\}=a t+\theta_1g_1=m$.

Inspect report $1$ first, stop on its high outcome, and otherwise use the common-reserve policy on reports $i\ge3$. Lemma~\ref{lem:binary:common} gives continuation utility $t$ and revenue $\E[(Z-t)^+]$. By~\eqref{eq:binary:blind-replacement}, the whole policy attains utility $m$, with expected payment ${p_1+a\E[(Z-t)^+]=C}$.
\end{proof}

\begin{lemma}[Making the second-highest-mean report free]
\label{lem:binary-free-second}
Set $p_2=0$, $p_1=v_1(h_2)$, and $p_i=\infty$ for $i\ge3$. Revenue is at least $D=\theta_2v_1(h_2)$.
\end{lemma}
\begin{proof}
Inspect the free report $2$ first. When $X_2=h_2$, all other blind means are at most $t\le h_2$, so report $1$ is worth exactly its fee $v_1(h_2)$. Seller-favorable ties induce purchase. This event has probability at least $\theta_2$, including coincident endpoints; other payments are nonnegative.
\end{proof}

\section{Proof of the revenue gap}
\label{app:commitment-proofs}

\subsection{The demand-integral bound}
\label{app:demand-integral}

\begin{lemma}[Uniform pricing and total information value]
\label{gen:uniformlog}
For $n\ge2$, we have $\OPT_{\unif}\ge\Delta/(1+\ln n)$.
\end{lemma}

\begin{proof}[Proof of Lemma~\ref{gen:uniformlog}]
Let $U(t)$ be optimal buyer utility when every inspection costs $t\ge0$,
and for $t>0$ let $Q(t)$ be the expected number of inspections under a
seller-favorable optimal response. Set $Q(0)=n$, implemented by complete free inspection. A fixed search policy $\pi$ has expected
selected value $v_\pi$ and expected inspection count $q_\pi\in[0,n]$.
Therefore
$U(t)=\sup_\pi\{v_\pi-tq_\pi\}$
is convex, nonincreasing, and $n$-Lipschitz. Free complete revelation
gives $U(0)=\mathbb E[\max_iX_i]$, and immediate stopping gives
$U(t)\ge\mu_*$. If $t>\Delta$, any policy that starts inspecting surely
has utility at most $\mathbb E[\max_iX_i]-t<\mu_*$. Thus $U(t)=\mu_*$
for all $t>\Delta$.

At every differentiability point $t>0$, any optimal policy's affine utility supports $U$ from below, so $-U'(t)=Q(t)$. Convexity gives differentiability almost everywhere, and the Lipschitz property gives absolute continuity. Thus
\begin{equation}
 \Delta=\int_0^\infty Q(t)\,dt.
 \label{gen:integraldemand}
\end{equation}
If $\Delta=0$, the claim is immediate, so suppose $\Delta>0$.

Write $R=\OPT_{\unif}$ and $\tau=\sup\{t:U(t)>\mu_*\}$. For $0<t<\tau$, stopping is strictly suboptimal, so $Q(t)\ge1$ and $R\ge t$. Taking the supremum gives $R\ge\tau$. For $t>\tau$, utility is constant and demand is zero. Together with $Q(t)\le\min\{n,R/t\}$, equation~\eqref{gen:integraldemand} gives
\[
\Delta\le\int_0^{R/n}n\,dt+\int_{R/n}^{R}\frac{R}{t}\,dt=R(1+\ln n).\qedhere
\]
\end{proof}

\subsection{Extraction through the first purchase}
\label{app:dynamic-proof}

\begin{proof}[Proof of Proposition~\ref{gen:dynamicexact}]
Under~\eqref{gen:dynamicfee}, every purchasing policy pays $\Delta$. Complete inspection gives utility $M-\Delta=\mu_*$, matching the blind payoff. No policy can do better: expected selected value is at most $M$, and a blind choice yields at most $\mu_*$. Seller-favorable ties therefore give revenue $\Delta$, attaining the bound in Lemma~\ref{lem:model:surplus}. After any first purchase, complete inspection is optimal because all further reports are free. This also covers $\Delta=0$.

For $\Delta>0$, discount the first fee to $(1-\eta)\Delta$, where $0<\eta<1$. Complete inspection then gives utility $\mu_*+\eta\Delta>\mu_*$. Every optimal response must purchase and pays $(1-\eta)\Delta$, regardless of later ties, proving the robustness statement in Section~\ref{sec:dynamic}.
\end{proof}

\subsection{The exponential lower bound}
\begin{proof}[Proof of Theorem~\ref{gen:adaptivegap}]
Proposition~\ref{gen:dynamicexact} gives $\OPT_{\dyn}=\Delta$, while Lemma~\ref{gen:uniformlog} gives
$\OPT_{\st}\ge\OPT_{\unif}\ge\Delta/(1+\ln n)$. This proves the upper bound.

For the lower bound, take independent $\operatorname{Exp}(1)$ values, with $\mu=1$ and $c=e^{-1}$. By Theorem~\ref{thm:iid:uniform}, it suffices to optimize a common price $p\in[0,c]$. For $p>0$, equation~\eqref{eq:iid:thresholds} gives $g(p)=-\ln p$ and $p=\ell(p)-1+e^{-\ell(p)}$. Writing $q=1-p$ and $z=1-e^{-\ell(p)}$, the revenue formula becomes $\mathcal R_n(p)=1-(1-p)^n-pz^{n-1}$. For $0<p<c$, we have $0<z\le q$, $z'(p)=(1-z)/z$, and $p=-\ln(1-z)-z=\int_0^z t/(1-t)\,dt$.
Consequently $p(1-z)\le z^2/2\le z^2$, and differentiation gives
\[
\begin{aligned}
\mathcal R_n'(p)
&=nq^{n-1}-z^{n-1}-p(n-1)(1-z)z^{n-3}\\
&\ge n\bigl(q^{n-1}-z^{n-1}\bigr)\ge0.
\end{aligned}
\]
Thus the optimum is at $p=c=e^{-1}$, with ${q=z=1-e^{-1}}$, giving
$\OPT_{\st}=1-(1-e^{-1})^{n-1}.$
Integrating the maximum's tail and substituting $y=1-e^{-x}$ gives $\E[\max_iX_i]=\int_0^1(1-y^n)/(1-y)\,dy=H_n$. Thus $\OPT_{\dyn}=H_n-1$, proving the lower bound. Its denominator tends to one and $H_n/\ln n\to1$, giving the stated limit.
\end{proof}

\end{document}